\documentclass{article}
\usepackage{spconf,amsmath,graphicx}
\usepackage{epsfig,amssymb}
\usepackage{subcaption}
\usepackage{multirow}
\usepackage[table,xcdraw]{xcolor}
\usepackage{hyperref}
\hypersetup{
    colorlinks=true,
    linkcolor=blue,
    urlcolor=blue,
    }

\newcommand{\vect}[1]{\boldsymbol{\mathrm{#1}}}

\newcommand{\norm}[1]{\left\lVert#1\right\rVert}

\title{STC SPEAKER RECOGNITION SYSTEM FOR THE NIST SRE 2021}

\name{ \begin{tabular}{c}Anastasia Avdeeva$^1$, Aleksei Gusev$^1$, Igor Korsunov$^1$, Alexander Kozlov$^1$, Galina Lavrentyeva$^1$, \\ Sergey Novoselov$^1$, Timur Pekhovsky$^1$, Andrey Shulipa$^2$, Alisa Vinogradova$^1$, \\ Vladimir Volokhov$^1$, Evgeny Smirnov$^1$, Vasily Galyuk$^1$ \end{tabular}}

\address{
$^1$ STC Ltd., St. Petersburg, Russia \\
$^2$ ITMO University, St. Petersburg, Russia \\
\small \tt {\{avdeeva-a, gusev-a, korsunov, kozlov-a, lavrentyeva, novoselov, tim, shulipa}, \\ \small \tt{gazizullina, volokhov, smirnov-e, galyuk\}@speechpro.com}}

\begin{document}
%\ninept
%
\maketitle
\begin{abstract}
This paper presents a description of STC Ltd. systems submitted to the NIST~2021 Speaker Recognition Evaluation for both fixed and open training conditions. These systems consists of a number of diverse subsystems based on using deep neural networks as feature extractors. 
During the NIST 2021 SRE challenge we focused on the training of the state-of-the-art deep speaker embeddings extractors like ResNets and ECAPA networks by using additive angular margin based loss functions. Additionally, inspired by the recent success of the wav2vec 2.0 features in automatic speech recognition we explored the effectiveness of this approach for the speaker verification filed. According to our observation the fine-tuning of the pretrained large wav2vec 2.0 model provides our best performing systems for open track condition. Our experiments with wav2vec 2.0 based extractors for the fixed condition showed that unsupervised autoregressive pretraining with Contrastive Predictive Coding loss opens the door to training powerful transformer-based extractors from raw speech signals.

For video modality we developed our best solution with RetinaFace face detector and deep ResNet face embeddings extractor trained on large face image datasets.

The final results for primary systems were obtained by different configurations of subsystems fusion on the score level followed by score calibration.
\end{abstract}
\begin{keywords}
Res\uppercase{n}et, speaker recognition, wav2vec.
\end{keywords}
\section{Introduction}
\label{sec:intro}
Today's state-of-the-art \cite{Zeinali2019, garcia2020magneto, gusev2020deep, lee2019nec, evalplan2021nist} speaker recognition systems are based on very deep convolutional neural networks (ResNets, ECAPAs, Extended TDNNs) which use log Mel Filter Bank features as input and are trained on large datasets using additive angular margin loss functions and different optimization strategies. The simple cosine or PLDA scoring are usually used as an extractors back-end. In our study we decided to follow this principles while developing systems for the SRE 21 Challenge. 

In contrast to the past NIST SREs \cite{sadjadi20172016, sadjadi20202019, matvejka202013} the key challenges provided by new NIST SRE 21 datasets \cite{evalplan2021nist} are multi-channel and multi-language speaker recognition based on audio-from video and telephone speech segments. Taking this into account the top performing systems should be well calibrated and robust for the cross-channel and same-channel microphone and telephone conditions. To this end we considered both 8 kHz and 16 kHz acoustic features to train different system.

Inspired by the success of wav2vec 2.0 in speech recognition tasks \cite{schneider2019wav2vec, baevski2020wav2vec} in our work we performed new study of wav2vec 2.0 model fine tuning for speaker recognition tasks. The experiments with wav2vec models were conducted for both fixed
and open track conditions.

% For the open track we used large multi-lingual wav2vec 2.0 model $XLSR\_53$ provided by facebook \cite{ott2019fairseq} on \href{https://github.com/pytorch/fairseq/tree/main/examples/wav2vec}{\textit{fairseq cite}} as a starting point of our fine-tuning. 

% For the fixed conditions we pretrained base wav2vec 2.0 model using Contrastive Predicting Coding \cite{oord2018representation} scheme on SRE 21 train set using different types of augmentations and cutting of non-speech segments. We used fairseq tools to do such pretraining.

It should be noted that wav2vec 2.0 models are powerfull transformer based models which take raw speech signals as input and incorporate multi-head attention mechanism on the deep layers.

During our investigations we found out that last classification layers of the extractors contain useful information for speaker verification. We explored some naive ideas of using this information by doing speaker verification on the classification layer output (cl-embeddings) with simple cosine similarity metric scoring. 

This paper presents the detailed description of the systems submitted by STC Ltd. to NIST SRE 2021 and its performance estimates obtained on the dev set.
\section{Speaker verification systems}
\label{sec:audio}

\subsection{Train datasets}
\label{sec:train dataset}

\textbf{Fixed-track train set.}
\label{Fix train datasets}
This set consists of data from
\begin{itemize}
    \item NIST SRE CTS Superset (LDC2021E08);
    \item 2016 NIST SRE Evaluation Set (LDC2019S20);
    \item concatenated VoxCeleb 1 and 2 datasets.
\end{itemize} 

For augmentation purposes we used standard Kaldi augmentation recipe (reverberation, babble, music and noise) with freely available MUSAN and simulated Room Impulse Response (RIR) datasets.
%with additional augmentation methods, described in section \ref{Open train datasets}. 
Additionally, we applied SpecAugment \cite{specaugment} technique and simulated codecs effects by different types of low-pass, high-pass, band-pass and band-stop filters. As well we simulated telephone and microphone channels with different types of Finite Impulse Responses (FIR), computed on 2021 NIST SRE Development Set.

In total, this training set contains 725,983 records from 14,271 speakers.

\textbf{Open-track train set.}
\label{Open train datasets}
This set was used for open-track systems training.
For building it we used a wide variety of different datasets containing telephone and microphone data from private datasets and from those available online:
\begin{itemize}
    \item Switchboard2 Phases 1, 2 and 3;
    \item Switchboard Cellular;
    \item Mixer 6 Speech;
    \item NIST SREs 2004 - 2010;
    \item NIST SRE 2018 (eval set);
    \item concatenated VoxCeleb 1 and 2;
    \item RusTelecom v2; %extended versions of the Russian speech subcorpus named RusTelecom v2 
    \item RusIVR corpus.
\end{itemize}
RusTelecom v2 is an extended versions of private Russian corpus of telephone speech, collected by call-centers in Russia. 
RusIVR is a private Russian corpus with telephone and media data, collected in various scenarios and recorded by different types of devices (telephone, headset, far-field microphone, etc).
In order to increase the amount and diversity of the training data, we used Kaldi standard recipe in addition to augmentations described in section \ref{Fix train datasets}.
In total, this training set contains 532,541 records from 33,466 speakers.

\textbf{Open-track tuning set.}
\label{Tune open train datasets}
This set is a subset of the \ref{Open train datasets} set and was used for tuning purposes only. 
Tuning set includes:
\begin{itemize}
    \item Mixer 6 Speech;
    \item NIST SREs 2004--2010, 2016;
    \item concatenated VoxCeleb 1 and 2;
    \item RusIVR.
\end{itemize}

Additionally, we preprocessed NIST SREs datasets: we fixed a number of speaker markup errors, discarded files with multiple speakers in one utterance, and added microphone data for some speakers. 
We also filtered out hard examples from the training dataset using Sub Center ArcFace technique \cite{arcface}. 

Standard Kaldi augmentation recipe was applied for this subset. In total, tuning set contains 336,724 records from 14,399 speakers.

\subsection{Feature extraction}

\textbf{16kHz features}.
We use Log  Mel-filter  bank  (MFB)  energies with the following extraction settings:
\begin{itemize}
\item frame-length -- 25 ms;
\item frame-shift -- 10 ms;
\item low frequency -- 20 Hz;
\item high frequency -- 7600 Hz;
\item number of mel bins -- 80.
\end{itemize}
After the features were extracted Mean Normalization (MN) over a 3-second sliding window was applied. 
In the fixed track we used Kaldi energy based VAD with energy threshold equal to 5.5. In open track U-net-based VAD was used instead.

\textbf{8kHz features}. 
We use Log  Mel-filter  bank  (MFB)  energies with the following extraction settings:
\begin{itemize}
\item frame-length -- 25 ms;
\item frame-shift -- 10 ms;
\item low frequency -- 20 Hz;
\item high frequency -- 3700 Hz;
\item number of mel bins -- 64.
\end{itemize}

Similarly to 16kHz data Mean Normalization and VAD was applied further.

\textbf{Raw audio signal processing.}
For our wav2vec based extractors we used raw 16 kHz audio. Kaldi-based energy vad was used for non-speech segment filtration. Additionally on-line augmentation scheme was used for raw audio samples with the following settings:
\begin{itemize}
    \item MUSAN additive noise with $p=0.25$;
    \item RIR convolution with $p=0.25$;
    \item Frequency masking with $p=0.25$;
    \item Time masking with $p=0.25$;
    \item Clipping Distortion with $p=0.25$.
\end{itemize}
Here $p$ is a probability of applying augmentation type for the sample in the training batch. All considered augmentations were applied in sequence. 
    
\subsection{Systems}
This section contains the description of all single systems used for final submission in fixed and open tracks. All of them contain corresponding suffix for clarity. 

During all stages of training and tuning processes AAM-Softmax loss was used with parameters $m$ and $s$ set to 0.35 and 32 respectively.
\subsubsection{Fixed track}
\textbf{ResNet34-16k-fixed}.
This extractor is based on the ResNet34 model with some modifications, as well as set to one stride in the first BasicBlock and changed to a simple Conv2D stem block
This is a ResNet34 model trained on 16kHz version of \ref{Fix train datasets}.

\textbf{ExtResNet34-8k-fixed}.
This is an extended Resnet34 model with bigger amount of filters on the frame level. It was trained on 8kHz version of \ref{Fix train datasets}. The model was initially trained for 15 epochs with 6 seconds speech segments using one cycle lr policy \cite{onecycle}. And then it was tuned for 3 epochs with crop size of 10 seconds and one epoch with 20 seconds.

\textbf{ExtResNet52-8k-fixed and ExtResNet52-16k-fixed}.
These models are two similar extended ResNet52 architectures constructed from BasicBlocks. We trained these models on 8kHz and 16kHz versions of \ref{Fix train datasets}. We have used only clear data and half of the augmented data, chosen at random to speed up experiments. These models were trained with scheme described above except last step with tuning on 20 seconds segments.  %During all stages we apply AAM-Softmax loss with parameters $m$ and $s$ set to 0.35 and 32 respectively.

\textbf{ResNet101-8k-fixed}.
This model is a modified version of the standard ResNet101 architecture with Basic Blocks instead of BottleneckBlocks. It was trained on the 8kHz version of \ref{Fix train datasets} for 20 epochs with a crop size of 5 seconds. The tuning procedure was performed twice for 3 epochs with 10 and 20 seconds segments length correspondingly.

\textbf{ResNet101-16k-fixed}.
\label{ResNet101_16k-fixed}
This model uses ResNet101 architecture with some modifications: set to one stride in the first BottleneckBlock and changed to a simple Conv2D stem block, which provides the basis for this extractor.
Model was trained on the 16kHz version of \ref{Fix train datasets} dataset with AMP in several stages with increasing crop size, loss margin and decreasing learning rate.

\textbf{ExtDResNet101-16k-fixed}.
This model is based on ResNet101\_8k-fixed architecture, which means that it also employs BasicBlocks. Additionally, it applies the model tweak from \cite{He_2019_CVPR} called ResNet-D: adding a $2\times2$ average pooling layer in the downsampling block with a stride of 2 before the convolution, whose stride is changed to 1. The model was trained on 16kHz data from \ref{Fix train datasets}. The first 20 epochs speech segments of 5 seconds length are used and then the model is tuned for 10 and 20 seconds segments.

\textbf{ECAPA-TDNN-fixed}.
Emphasized Channel Attention, Propagation and Aggregation in TDNN (ECAPA-TDNN), newly proposed in \cite{desplanques2020ecapa}, is a modification of the standard Time Delay Neural Network (TDNN) architecture, containing Squeeze-Excitation (SE) blocks and Res2Net modules at the frame level and attentive statistic pooling (ASP) \cite{Okabe2018} instead of the usual statistic pooling. We use our implementation of ECAPA-TDNN architecture with the following parameters: the number of SE-Res2Net Blocks is set to 4 with dilation values 2,3,4,5 to blocks; the number of filters in the convolutional frame layers C is set to 2048; the number of filters in the bottleneck set to 1536 of the SE-Res2Net Block; ASP is used; embedding layer size is set to 1024; simple Conv1D with 2048 filters is used like a stem block.  Model training is performed on 16kHz version of \ref{Fix train datasets} dataset with AMP in several stages.

\textbf{ResNet101-16k-fixed \& ECAPA-TDNN-fixed.}
This system is the embedding level fusion of the ResNet101-16k-fixed and ECAPA-TDNN-fixed systems. However, instead of standard embeddings, the class posteriors logit embeddings were used here. In more details they are described in section \ref{sec:cl_emb}.

\begin{figure}[t!]
\centering
\includegraphics[scale=0.85]{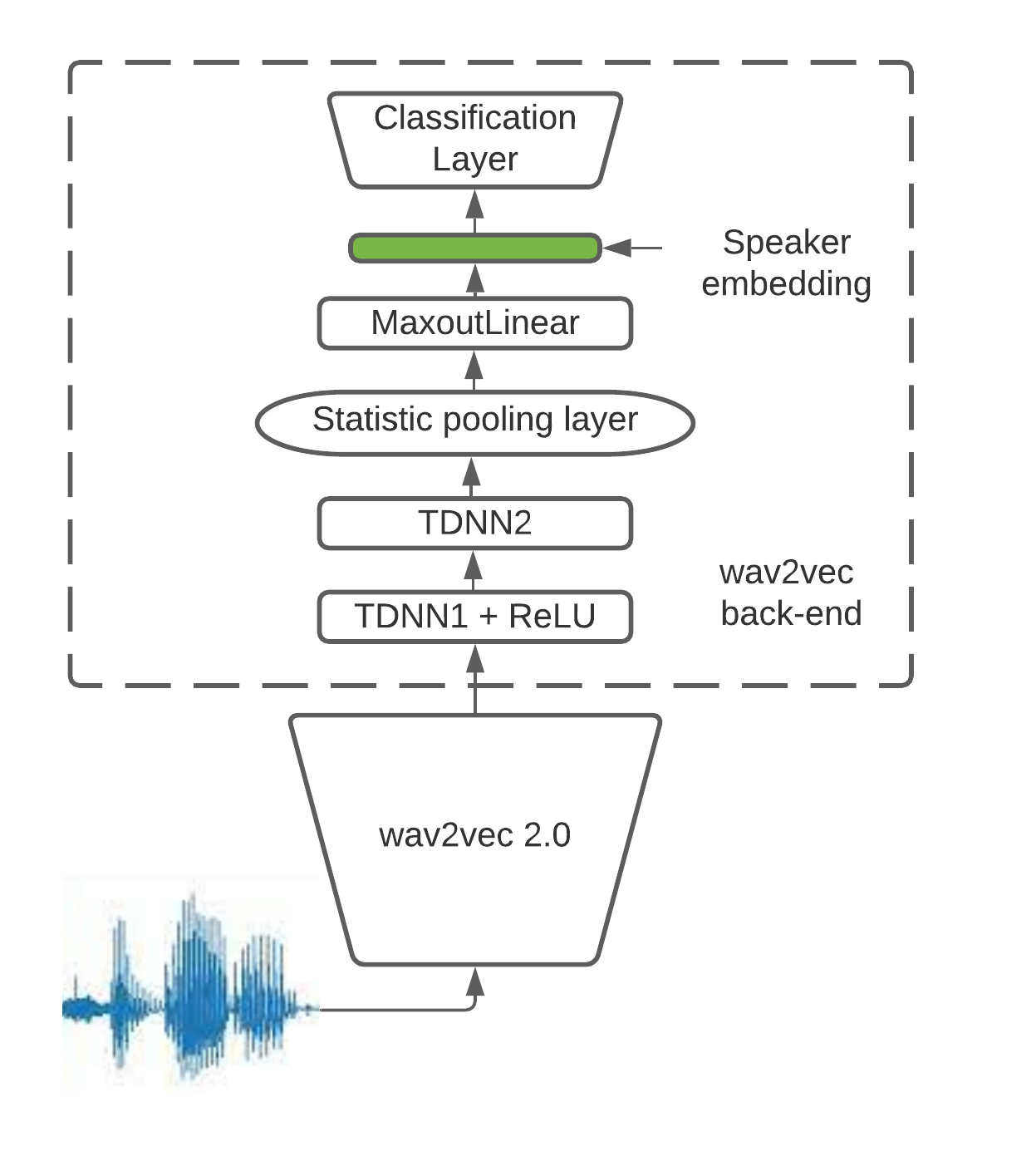}
\caption{Wav2vec 2.0 based speaker embeddings extractor}
\label{fig:w2v_arch}
\end{figure}

\textbf{Wav2Vec-fixed}. 
For the fixed conditions we pretrained base wav2vec 2.0 \cite{baevski2020wav2vec} model using Contrastive Predicting Coding \cite{oord2018representation} scheme on SRE 21 train set with different augmentations types  and cutting of non-speech segments. 
We used fairseq toolkit \cite{ott2019fairseq} \footnote{\href{https://github.com/pytorch/fairseq/tree/main/examples/wav2vec}{https://github.com/pytorch/fairseq/tree/main/examples/wav2vec}} for its pretraining. 

For training we used a part of fixed dataset - CTS\_Superset and Voxceleb 1, 2 with different sets of augmentations similar to described above in \ref{sec:train dataset}. We also applied Kaldi energy-based VAD for cutting long pauses in CTS\_Superset. Model was pretrained for 7 epoch and then used as a starting point for our finetuning. 
    
The main scheme of wav2vec 2.0 based speaker embeddings extractor is represented on Figure \ref{fig:w2v_arch}. As an effective wav2vec 2.0 back-end we applied two TDNN layers ( the  1st with ReLU activation), statistic pooling layer to pool time series to single vector, maxout linear layer \cite{novoselov2018deep, gusev2020deep} to get speaker embedding. We used AAM-Softmax based linear classification layer to fine-tune the extractor. 
%We should note that one can use wav2vec output as a statistic pooling input directly, but our intuition here is as follows.
In principle, one can pass output of the wav2vec directly to the statistics pooling layer. 
However, we find out that we can achive better results if we pass them through the sequence of TDNN layers.
%However, we had the knowledge that unsupervised wav2vec model pretraining leads to good speech specific information generalization on the top layers of the autoregressive model. 
The role of TDNN layers is to prefilter speaker specific information and to "prepare" wav2vec output time series for statistical pooling. According to our observations this approach let us achieve better results than direct statistical pooling of the wav2vec outputs.
The TDNN blocks utilise contex 1 of the input features and have 2048 dimension output. The obtained final speaker embedding size was 512. 
Additional note is that wav2vec part of the extractor could be freezed while tuning for downstream speaker recognition task. We observed that in this scenario the results can also be very good, but fine-tuning the whole extractors provides additional performance gains for speaker recognition system.

\textbf{SWIN-16k-fixed}.
This model is based on a %hierarchical transformer architecture
Shifted Windows Transformer architecture, proposed in \cite{swin} as an adaptation of standard transformers from NLP to computer vision task.

In our experiments we used the following architecture settings: 

\begin{itemize}
    \item image size: [80, 512]
    \item patch size: [2, 2]
    \item embedding dim of Swin Transformer Block 72
    \item model depths: [2, 4, 8, 4]
    \item number of attention heads for Swin Transformer Block: [3, 3, 3, 3]
    \item window size: 8.
\end{itemize}

We have added the statistics pooling layer and linear projection layer on top of the original model to obtain speaker embeddings. The model was trained  on 16kHz version of \ref{Fix train datasets} dataset with AMP in several stages with increasing crop size and loss margin and decreasing learning rate.

\subsubsection{Open}
For the final systems in open audio and audio-visual tracks we used systems from the fixed tracks described above as well as ones trained on the extended datasets.

\textbf{ECAPA-TDNN-open}.
This model uses ECAPA-TDNN architecture \cite{desplanques2020ecapa} with the following parameters:
    \begin{itemize}
        \item 4 SE-Res2Net Blocks with dilation values 2,3,4,5
        \item 1024 filters in convolution frame layers C to match the number of filters in the bottleneck of the SE-Res2Net Block
        \item adaptive statistics pooling
        \item embedding size 512
    \end{itemize}
    
We changed stem block to the stack of 4 Conv2D-BatchNorm2D-ReLU sequences with kernel size 3 and 32 filters for all convolution layers except the last one that used 1024 filters. Model was trained on 8kHz version of \ref{Open train datasets} dataset. The training process was performed in several stages with simultaneous increasing of the crop size (from 5 to 12 seconds) and loss margin and learning rate decreasing . After that the classification layer was reinitialised and the model was fine-tuned on \ref{Tune open train datasets} dataset.

\textbf{ResNet101-8k-open}.
This model is based on ResNet101 with some modifications: maxout activation function for the embedding layer, stride of one in BoottleneckBlock and simple Conv2D layer in place of stem block. Model was trained on 8kHz version of \ref{Open train datasets} dataset and tuned on Open-track tune set \ref{Tune open train datasets}  similarly to ECAPA-TDNN-open.

\textbf{ResNet101-8k-open + ECAPA-TDNN-open} ~~~~~~~~~~~~and \\ \textbf{ResNet101-8k-open + ECAPA-TDNN-open filtered}. 
Both of these systems were prepared similarly to the systems for the fixed track: ResNet101-8k-open and ECAPA-TDNN-open were fused together on the cl-embeddings level (see section \ref{sec:cl_emb}).

We eliminated less informative speaker classes in ResNet101-8k-open + ECAPA-TDNN-open filtered after the fusion to reduce the size of cl-embeddings.
This technique produces better results on the dev set (Table \ref{tab:single_open}).

\textbf{Wav2vec-(I,II,III,IV)-open}. 
These models were based on wav2vec 2.0 large architecture.
For the open track we used large multi-lingual wav2vec 2.0 model $XLSR\_53$ provided by facebook \cite{ott2019fairseq} on \href{https://github.com/pytorch/fairseq/tree/main/examples/wav2vec}{\textit{fairseq cite}} as a starting point for the fine-tuning. We used the same wav2vec Back-End as in our fix condition model. But the Wav2vec-(I,II,III)-open were trained using (8 sec, 12 sec, 14 sec) speech chunks durations respectively with different training strategies on \ref{Open train datasets} dataset. Wav2vec-IV-open was fine-tuned on \ref{Fix train datasets} dataset using 14 sec speech chunks duration.  

\subsection{Class posteriors logit embeddings}
\label{sec:cl_emb}
During our investigation we observed that the extractors classification layer outputs (namely class posteriors logit embeddings, or cl-embeddings) can be more informative than conventional pre-last layer embeddings. We realised that last classification layers obtained in closed classification task discriminative training contain useful information for the open task speaker discrimination. We explored some naive ideas of using this information by doing speaker verification on the classification layer output  with simple cosine similarity metric scoring. For instance, the results of our experiments with cl-embeddings for two type of extractors  ResNet101-16k-fix and ECAPA-TDNN-fix are presented in Table \ref{tab:my-table}. Here cl-embeddings outperform conventional embeddings in terms of both the EER and minDCF. Moreover one can notice that cl-embeddings subspace allows to use effective and simple embeddings fusion procedure like weighted cl-embeddings sum. But this is possible if the extractors had the same training speakers classes or on intersected classes. From the results of Table \ref{tab:my-table} you can see such type of fusion is superior then simple score level fusion of the systems.

We use cl-embedding based fusion for ResNet101-16k-fix \& ECAPA-TDNN-fix models  and consider it further as single systems in section \ref{sec:submission}.

\subsection{Scoring}
We used Cosine similarity to distinguish speaker embeddings:
\begin{equation}
\mathcal{S(\vect{x_1},\vect{x_2})} = \dfrac{\vect{x_1}^T\vect{x_2}}{{\norm{\vect{x_1}}}{\norm{\vect{x_2}}}},
\end{equation}
where $(\vect{x_1}, \vect{x_2})$ are speaker embedding vectors.

\subsection{Score normalization}
For all systems except of those based on wav2vec, adaptive scoring normalization technique (adaptive s-norm) from \cite{CVDFKCL2017} is used. Here the normalized score for a pair $(\vect{x_1},\vect{x_2})$ can be estimated as follows:
\begin{equation}
\label{eq:score_norm}
\mathcal{\hat{S}(\vect{x_1},\vect{x_2})} = \frac{\mathcal{S(\vect{x_1},\vect{x_2})}-\mu_1}{\sigma_1}+\frac{\mathcal{S(\vect{x_1},\vect{x_2})}-\mu_2}{\sigma_2},
\end{equation}
where the mean $\mu_1$ and standard deviation $\sigma_1$ are calculated by matching $\vect{x_1}$ against impostor cohort and similarly for $\mu_2$ and $\sigma_2$. A set of the $\textit{n}$ best scoring impostors are selected for each embedding pair when means and standard deviations are calculated.
\begin{figure}[t]

\begin{subfigure}{\linewidth}
  \centering
  \includegraphics[width=\linewidth]{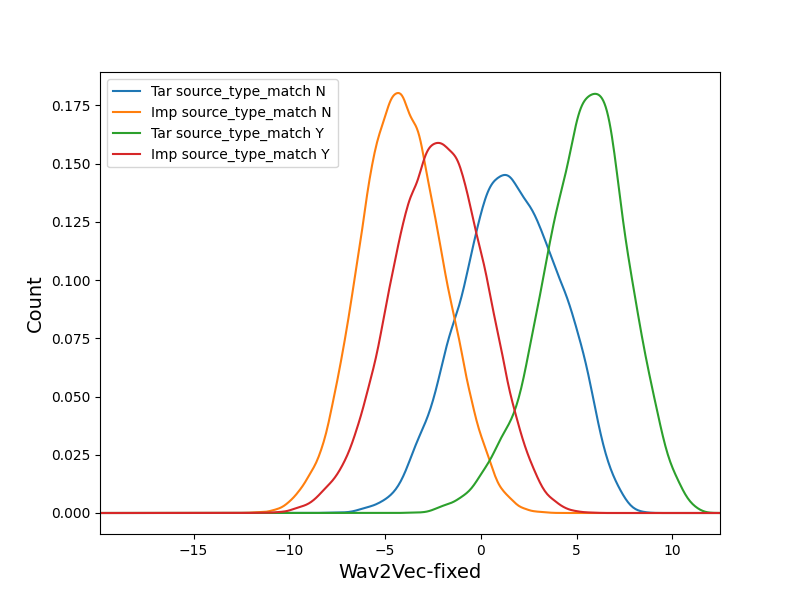}
\end{subfigure}
\begin{subfigure}{\linewidth}
  \centering
  \includegraphics[width=\linewidth]{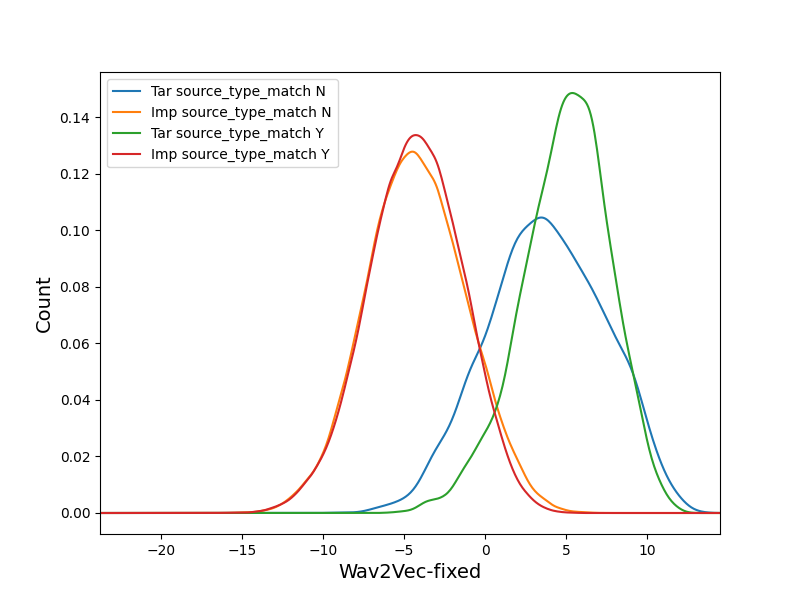}
\end{subfigure}
\caption{Comparison of target and impostor distributions with and without source type match obtained for Wav2Vec-fixed system: (a) no normalization applied, (b) channel normalization applied}
\label{img:dist}
\end{figure}
Additionally, we analysed the scores distribution for our single systems for dev set with and without source type match. The obtained results in Figure \ref{img:dist} confirm significant gap between the corresponding distributions. This led us to the idea of channel compensation.

For all systems scores we apply channel normalization technique, where the normalized score for a pair $(\vect{x_1},\vect{x_2})$ can be estimated as follows:
\begin{equation}
\label{eq:score_norm2}
\mathcal{\hat{S}(\vect{x_1},\vect{x_2})} = \frac{\mathcal{S(\vect{x_1},\vect{x_2})}-\mu_{ch}}{\sigma_{ch}},
\end{equation}
where mean $\mu_{ch}$ and standard deviation $\sigma_{ch}$ are calculated for each pair of source type matching (tel-tel, mic-mic, tel-mic and mic-tel), obtained from source files headers and applied according to the the source type of $(\vect{x_1},\vect{x_2})$.

For those systems that used both normalization approaches, adaptive s-norm is applied before channel normalization.

Tables \ref{tab:single_fixed} and \ref{tab:single_open} demonstrate quality estimates in terms of EER, minDCF and ActDCF obtained on the dev set for each single system in its better configuration for fix and open tracks correspondingly.

\begin{table*}[t]
\centering
\caption{Results of cl-embeddings, scores normalization and fusion for ResNet101-16k-fix and ECAPA-TDNN-fix systems for the audio track. Here cl-emb is the flag of using class posterior logits embeddings, ch-norm is channel scores normalization flag, s-norm -- adaptive scores normalization flag respectively. The metrics were computed by NIST SRE 21 scoring tool}
\vspace{2mm}
\label{tab:my-table}
\centerline{
\begin{tabular}{|c|l|l|l|l|l|}
\hline
\rowcolor[HTML]{FFFFC7} 
\multicolumn{1}{|l|}{\cellcolor[HTML]{FFFFC7}\textbf{System   name}}                                                & \textbf{cl-emb} & \textbf{s-norm} & \textbf{ch-norm} & \textbf{EER, \%} & \textbf{minDCF21} \\ \hline
 & - & - & -  & 8.04             & 0.408             \\ \cline{2-6} 
 & - & \checkmark & \checkmark    & 4.65             & 0.316             \\ \cline{2-6} 
 & - & - & \checkmark & 5.28             & 0.36              \\ \cline{2-6} 
 & - & \checkmark & - & 6.88             & 0.412             \\ \cline{2-6} 
 & \checkmark & - & - & 6.93             & 0.35              \\ \cline{2-6} 
 & \checkmark & \checkmark & \checkmark & \textbf{4.40}              & \textbf{0.280}              \\ \cline{2-6} 
 & \checkmark  & - & \checkmark             & 5.06             & 0.309             \\ \cline{2-6} 
\multirow{-8}{*}{ResNet101-16k-fixed}                                                                     & \checkmark & \checkmark & -  & 5.94             & 0.336             \\ \hline
 & - & - & - & 10.47                & 0.499             \\ \cline{2-6} 
 & - & \checkmark & \checkmark & 7.09             & 0.451               \\ \cline{2-6} 
 & - & - & \checkmark & 8.05             & 0.455              \\ \cline{2-6} 
 & - & \checkmark & - & 10.34            & 0.593             \\ \cline{2-6} 
 & \checkmark  & - & - & 8.26             & 0.411             \\ \cline{2-6} 
 & \checkmark & \checkmark & \checkmark & \textbf{5.94}             & \textbf{0.352}             \\ \cline{2-6} 
  & \checkmark & - & \checkmark             & 6.50             & 0.376             \\ \cline{2-6} 
\multirow{-8}{*}{ECAPA-TDNN-fixed}                                                                                    & \checkmark            & \checkmark            & -            & 8.07             & 0.486              \\ \hline
\multicolumn{1}{|l|}{\begin{tabular}[c]{@{}l@{}}ResNet101-16k-fixed +\\  ECAPA-TDNN-fixed \\ (scores fusion)\end{tabular}}     & \checkmark            & \checkmark            & \checkmark             & 3.97             & 0.253             \\ \hline
\multicolumn{1}{|l|}{\begin{tabular}[c]{@{}l@{}}ResNet101-16k-fixed +\\  ECAPA-TDNN-fixed \\ (embeddings fusion)\end{tabular}} & \checkmark            & \checkmark            & \checkmark             & \textbf{3.50}             & \textbf{0.230}             \\ \hline
\end{tabular}}
\end{table*}

\begin{table}[]
\centering
\caption{Fixed audio track single systems}
\label{tab:single_fixed}
\begin{tabular}{|l|c|c|c|}
\hline
\rowcolor[HTML]{FFFFC7} 
\textbf{System} & \textbf{EER} & \textbf{minDCF} & \textbf{actDCF} \\ \hline
ResNet34-16k-fixed & 6.1 & 0.377 & 0.39 \\ \hline
\begin{tabular}[c]{@{}l@{}} ResNet101-16k-fixed \\+ ECAPA-TDNN-fixed \end{tabular}& 3.5 & 0.23 & 0.233 \\ \hline
Wav2Vec-fixed & 7.57 & 0.426 & 0.45 \\ \hline
ResNet101-8k-fixed & 4.24 & 0.322 & 0.333 \\ \hline
ExtDResNet101-16k-fixed & 4.68 & 0.255 & 0.271 \\ \hline
SWIN-16k-fixed & 6.72 & 0.443 & 0.465 \\ \hline
ExtResNet34-8k-fixed & 3.96 & 0.301 & 0.305 \\ \hline
ExtResNet52-8k-fixed & 3.67 & 0.266 & 0.278 \\ \hline
ExtResNet52-16k-fixed & 3.65 & 0.28 & 0.282 \\ \hline
\end{tabular}
\end{table}

\begin{table}[]
\centering
\caption{Open audio track single systems}
\label{tab:single_open}
\begin{tabular}{|l|l|l|l|}
\hline
\rowcolor[HTML]{FFFFC7} 
\textbf{System} & \textbf{EER} & \textbf{minDCF} & \textbf{actDCF} \\ \hline
ResNet34-16k-fixed & 6.56 & 0.377 & 0.387 \\ \hline
ExtResNet34-8k-fixed & 4.28 & 0.312 & 0.318 \\ \hline
ExtResNet52-8k-fixed & 3.95 & 0.283 & 0.287 \\ \hline
ExtResNet52-16k-fixed & 4.09 & 0.287 & 0.292 \\ \hline
ResNet101-8k-fixed & 4.16 & 0.298 & 0.303 \\ \hline
ExtDResNet101-16k-fixed & 4.7 & 0.265 & 0.277 \\ \hline
\begin{tabular}[c]{@{}l@{}}ResNet101-16k-fixed \\+ ECAPA-TDNN-fixed \end{tabular} & 6.69 & 0.426 & 0.435 \\ \hline
\begin{tabular}[c]{@{}l@{}}ResNet101-8k-open \\+ ECAPA-TDNN-open \end{tabular}& 2.88 & 0.209 & 0.215 \\ \hline
\begin{tabular}[c]{@{}l@{}}ResNet101-8k-open \\+ ECAPA-TDNN-open-filtered  \end{tabular}& 2.94 & 0.2 & 0.207 \\ \hline
Wav2vec-I-open & 4.49 & 0.27 & 0.276 \\ \hline
Wav2vec-II-open & 2.88 & 0.246 & 0.258 \\ \hline
Wav2vec-III-open & 3.68 & 0.2 & 0.203 \\ \hline
Wav2vec-IV-open & 3.4 & 0.233 & 0.247 \\ \hline
SWIN-16k-fixed & 6.91 & 0.446 & 0.459 \\ \hline

r101-l2\_ecapa-l2\_class\_fusion & 3.32 & 0.257 & 0.271 \\ \hline
\end{tabular}
\end{table}

\section{Face verification systems}
\label{sec:video}

We used a standard pipeline to solve the face verification problem: face detection, preprocessing of facial crops, embedding extraction, and scoring.

Firstly, a digital image of enroll or many frames of test video were processed using a face detector. The outputs of the face detector are the coordinates of the bounding box and the coordinates of the five facial landmarks. Secondly, we used the coordinates of the bounding box to create facial crops. The facial crops were aligned using the five facial landmarks, and then resized to $112\times112$ pixels, and normalized. Thirdly, we extracted facial embeddings for enroll and test crops and, fourthly, we performed a scoring between enroll and test embeddings for each trial in development, and evaluation protocols.

\subsection{Train datasets}

We used the existing public face detector model based on RetinaFace \cite{RetinaFace} and focused on the high quality facial embedding extractor training. For this purpose we used the following databases: MS-Celeb-1M \cite{guo2016ms}, VGGFace2 \cite{cao2018vggface2}, TrillionPairs-Asians  \cite{trillionpairs}, DFW2018 \cite{kushwaha2018disguised} and our proprietary database.

\subsection{Systems}

\subsubsection{Preprocessing stage}

As described above, we used \textbf{RetinaFace} as a face detector. Since enrolls are described by an image, we used the face detector once to detect enroll crop of face. Tests are described by an video. We used FFmpeg \cite{ffmpeg} to extract every ten frames of test video and processed each test frame using RetinaFace. We did not use any face tracking algorithms.

\begin{figure}
\centering
\includegraphics[width=.8\linewidth]{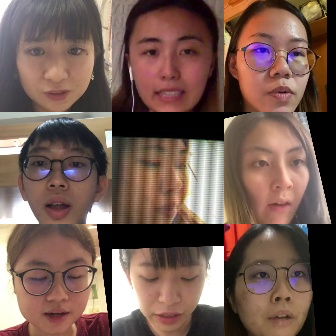}
\caption{Examples of facial crops ($112\times112$ pixels) after preprocessing stage from NIST SRE 2021 development set}
\label{fig:facial_crops}
\end{figure}

As noted in the original paper \cite{RetinaFace}, RetinaFace is a single-stage pixel-wise face localisation method, which employs extra-supervised and self-supervised multi-task learning in parallel with the existing box classification and regression branches. Each positive anchor outputs a face score, a face box, five facial landmarks, and dense 3D face vertices projected on the image plane. We used the coordinates of the bounding box to create facial crops and we used five facial landmarks, located in the area of the eyes, the tip of the nose, and the tips of the lips, to align facial crops. All facial crops have been resized to $112\times112$ pixels. We illustrated the final facial crops after preprocessing of test frames from NIST SRE 2021 development set on Figure~\ref{fig:facial_crops}.

An important point to note in relation to visual data from NIST SRE 2021 development and evaluation set is header information in test video. The header of each test video contains rotation parameter. The rotation parameter determines the orientation of the frames from the test video. We read the rotation parameter to correctly orient test frames before performing the face detection procedure. We believe this parameter is important for the correct work of the preprocessing stage.

%Unfortunately, this parameter is not correct for all test videos and our preprocessing stage returns incorrect results of the description of the face crops. We would like to highlight this information for the participants and organizers of the challenge, because we think that with the correct rotation parameter the participant facial biometrics systems will work even better.

\subsubsection{Facial embedding extractors}

We trained two facial embedding extractors for the NIST SRE 2021 challenge. First embedding extractor was based on \textbf{ResNet101-IR-SE-AN} neural network architecture, where IR means inverted residual \cite{DGZ2019}, SE means squeeze-and-excitation \cite{HCS2018}, and AN means attentive normalization \cite{AN}. The embedding extractor was trained using MS-Celeb-1M, VGGFace2, TrillionPairs-Asians, DFW2018 datasets. 

Second embedding extractor was based on \textbf{ResNet100-PM} neural network architecture, where PM means prototype memory \cite{SGG2021}. Prototype memory is a face representation learning model, which alleviates "prototype obsolescence" problem (classifier weights (prototypes) of the rarely sampled classes, receive too scarce gradients and become outdated and detached from the current encoder state, resulting in an incorrect training signals) and allows training on a dataset of any size. We trained the ResNet100-PM embedding extractor using our proprietary database.

\subsubsection{Scoring methods}
ResNet101-IR-SE-AN's embeddings and ResNet100-PM's embeddings are normalized by length and can be distinguished by cosine similarity. Each trial in development and evaluation protocols contains a enroll and several test embeddings for visual modality. We used different scoring methods to compare enroll and test facial embeddings:

\begin{itemize}
\item calculating of maximum score (MS) between enroll embedding and test embeddings for each trial;
\item calculating of score between enroll embedding and average test embedding (ATE) for each trial;
\item calculating of score between enroll embedding and weighted average test embedding (WATE) for each trial (we don't use test embeddings with small recognizability score to compute weighted average test embedding);
\item calculating of score between enroll embedding and test embedding with maximum recognizability score (MRS) for each trial.
\end{itemize}

Face detector can detect faces that cannot be recognized, no matter how capable the recognition system is. Recognizability, a latent variable, can therefore be factored into the design and implementation of face recognition systems. We implemented a measure of recognizability of a face image using idea from \cite{RecScore}. We compute a cosine distance to each test embeddings from an embedding of "unrecognizable identity" as a measure of recognizability. Final score of recognizability to each test embeddings was computed using the following mathematical expression:
\begin{equation}
\mathcal{RS(\vect{x_t},\vect{x_{ui}})} = \dfrac{\left(\left(1 -  \dfrac{\vect{x_t}^T\vect{x_{ui}}}{{\norm{\vect{x_t}}}{\norm{\vect{x_{ui}}}}}\right) - 0.35\right)}{0.89},
\end{equation}
where $\vect{x_t}$ is a test embedding and $\vect{x_{ui}}$ is an embedding of "unrecognizable identity". The values 0.35 and 0.89 in the expression above was chosen empirically to create a dynamic range for $\mathcal{RS}$ between 0 and 1. We computed embeddings of "unrecognizable identity" to each our embedding extractors using unrecognizable images from WIDER FACE dataset \cite{wider_face}. We tried to use recognizability score to compute weighted average test embedding and to search the best test embedding for each trial.

\begin{table}[]
\centering
\caption{Visual track single systems}
\label{tab:visual}
\resizebox{\columnwidth}{!}{
\begin{tabular}{|l|c|c|c|}
\hline
\rowcolor[HTML]{FFFFC7} 
\textbf{System} & \textbf{EER} & \textbf{minDCF} & \textbf{actDCF} \\ \hline
ResNet101-IR-SE-AN + MS   & \textbf{0.17} & \textbf{0.011} & \textbf{0.013} \\ \hline
ResNet101-IR-SE-AN + ATE  & 0.99 & 0.018 & - \\ \hline
ResNet101-IR-SE-AN + WATE & \textbf{0.22} & \textbf{0.010} & \textbf{0.013} \\ \hline
ResNet101-IR-SE-AN + MRS  & 1.83 & 0.046 & - \\ \hline
ResNet100-PM + MS         & 1.32 & 0.013 & 0.018 \\ \hline
ResNet100-PM + ATE        & 1.82 & 0.018 & - \\ \hline
ResNet100-PM + WATE       & 1.66 & 0.017 & - \\ \hline
ResNet100-PM + MRS        & 2.94 & 0.046 & - \\ \hline
\end{tabular}
}
\end{table}

\begin{table}[]
\centering
\caption{Audio-visual track single systems (only visual)}
\label{tab:audio_visual}
\resizebox{\columnwidth}{!}{
\begin{tabular}{|l|c|c|c|}
\hline
\rowcolor[HTML]{FFFFC7} 
\textbf{System} & \textbf{EER} & \textbf{minDCF} & \textbf{actDCF} \\ \hline
ResNet101-IR-SE-AN + MS   & \textbf{0.10} & \textbf{0.007} & - \\ \hline
ResNet101-IR-SE-AN + ATE  & 0.60 & 0.011 & - \\ \hline
ResNet101-IR-SE-AN + WATE & \textbf{0.23} & \textbf{0.006} & - \\ \hline
ResNet101-IR-SE-AN + MRS  & 2.40 & 0.055 & - \\ \hline
ResNet100-PM + MS         & 0.80 & 0.008 & - \\ \hline
ResNet100-PM + ATE        & 1.10 & 0.011 & - \\ \hline
ResNet100-PM + WATE       & 1.00 & 0.010 & - \\ \hline
ResNet100-PM + MRS        & 2.90 & 0.054 & - \\ \hline
\end{tabular}
}
\end{table}

\begin{table*}[]
\caption{Results of the systems prepared for the NIST SRE 21 challenge on the development set}
\label{tab:submission}
\vspace{4mm}
\resizebox{\textwidth}{!}{
\begin{tabular}{|l|l|c|c|c|l|c|c|c|}
\hline
\rowcolor[HTML]{FFFFC7} 
 & Fixed audio track & EER & minDCF & actDCF & Open audio track & EER & minDCF & actDCF \\ \hline
\rotatebox[origin=c]{90}{Primary} & \begin{tabular}[c]{@{}l@{}}
ResNet101-16k-fixed \\ +  ECAPA-TDNN-fixe\\ Wav2Vec-fixed\\ ResNet101-8k-fixed\\ ExtDResNet101-16k-fixed\\  ExtResNet52-16k-fixed\end{tabular} & 3.09 & 0.216 & 0.220 & \begin{tabular}[c]{@{}l@{}}Wav2vec-III-open \\ ResNet101-8k-open \\ + ECAPA-TDNN-open filtered \\ SWIN16k-fixed \\ ResNet101-16k-fixed \\ +  ECAPA-TDNN-fixed \\ Wav2vec-IV-open \\ Wav2vec-I-open \\ ExtResNet52-16k-fixed \\ ResNet34-16k-fixed \\ ExtDResNet101-16k-fixed \\ ExtResNet34-8k-fixed \\ ExtResNet52-8k-fixed\end{tabular} & \multicolumn{1}{c|}{1.97} & 0.151 & 0.153 \\ \hline
\rotatebox[origin=c]{90}{Contrastive} & \begin{tabular}[c]{@{}l@{}}ResNet101-16k-fix \\ + ECAPA-TDNN-fix\end{tabular} & 3.49 & 0.230 & 0.233 & \begin{tabular}[c]{@{}l@{}}Wav2vec-III-open\\ ResNet101-8k-open \\ + ECAPA-TDNN-open filtered\end{tabular} & 2.38 & 0.156 & 0.161 \\ \hline
\rotatebox[origin=c]{90}{Single} & ExtResNet52-8k-fixed & 3.67 & 0.266 & 0.278 & Wav2vec-III-open & 2.24 & 0.207 & 0.212 \\ \hline
\rowcolor[HTML]{FFFFC7} 
 & Fixed visual track & EER & minDCF & actDCF & Open visual track & EER & minDCF & actDCF \\ \hline
\rotatebox[origin=c]{90}{Primary} & ResNet101-IR-SE-AN + MS & 0.17  & 0.011 & 0.013 & ResNet101-IR-SE-AN + MS & 0.17  & 0.011 & 0.013 \\ \hline
\rotatebox[origin=c]{90}{Contrastive} &
\begin{tabular}[c]{@{}l@{}} ResNet101-IR-SE-AN + MS \\ ResNet101-IR-SE-AN + WATE \end{tabular} & 0.17 & 0.010 & 0.012 & ResNet100-PM + MS & 1.32 & 0.013 & 0.018 \\ \hline
\rotatebox[origin=c]{90}{Single} & ResNet101-IR-SE-AN + MS & 0.17  & 0.011 & 0.013 & ResNet101-IR-SE-AN + WATE & 0.22 & 0.010 & 0.013 \\ \hline
\rowcolor[HTML]{FFFFC7} 
 & Fixed audio-visual track & EER & minDCF & actDCF & Open audio-visual track & EER & minDCF & actDCF \\ \hline
\rotatebox[origin=c]{90}{Primary} & \begin{tabular}[c]{@{}l@{}} \textit{audio:} \\ ResNet101-16k-fixed \\ + ECAPA-TDNN-fixed\\ Wav2Vec-fixed\\ ResNet101-8k-fixed\\ ExtDResNet101-16k-fixed\\  ExtResNet52-16k-fixed \\ \textit{video}: \\ ResNet101-IR-SE-AN + MS\end{tabular} & 0.10 & 0.002 & 0.003 & \begin{tabular}[c]{@{}l@{}} \textit{audio}: \\ Wav2vec-III-open \\ ResNet101-8k-open \\ + ECAPA-TDNN-open filtered \\ SWIN16k-fixed \\ ResNet101-16k-fixed \\ +  ECAPA-TDNN-fixed \\ Wav2vec-IV-open \\ Wav2vec-I-open \\ ExtResNet52-16k-fixed \\ ResNet34-16k-fixed \\ ExtDResNet101-16k-fixed \\ ExtResNet34-8k-fixed \\ ExtResNet52-8k-fixed \\ \textit{video:} \\ ResNet101-IR-SE-AN + MS\end{tabular} & 0.10 & 0.001 & 0.002 \\ \hline
\rotatebox[origin=c]{90}{Contrastive} & \begin{tabular}[c]{@{}l@{}} \textit{audio:} \\ ResNet101-16k-fix \\+ ECAPA-TDNN-fix1\_NIST\\ \textit{video:} ResNet101-IR-SE-AN + MS\end{tabular} & 0.10 & 0.002 & 0.003 & \begin{tabular}[c]{@{}l@{}}\textit{audio:} \\Wav2vec-III-open\\ ResNet101-8k-open \\ + ECAPA-TDNN-open filtered \\ \textit{video:} \\ ResNet101-IR-SE-AN + MS\end{tabular} & 0.10 & 0.002 & 0.003 \\ \hline
\rotatebox[origin=c]{90}{Single} & \begin{tabular}[c]{@{}l@{}}\textit{audio:} \\ ExtDResNet101-16k-fixed\\ \textit{video:} \\ ResNet101-IR-SE-AN + MS\end{tabular} & 0.13 & 0.003 & 0.003 & \begin{tabular}[c]{@{}l@{}}\textit{audio:} \\ Wav2vec-III-open\\ \textit{video:} \\ ResNet101-IR-SE-AN + MS\end{tabular} & 0.10 & 0.003 & 0.004 \\ \hline
\end{tabular}
}
\end{table*}

Numerical results for NIST SRE 2021 (development set) visual and audio-visual tracks (only visual part) without performing the calibration procedure are represented in Table~\ref{tab:visual} and Table~\ref{tab:audio_visual}. The results show that ResNet101-IR-SE-AN extractor significantly outperforms ResNet100-PM extractor. We think that this is due to the fact that TrillionPairs-Asians database was used to train ResNet101-IR-SE-AN extractor. This database contains many Asian faces as well as NIST SRE 2021 development an evaluation set. It is showed from Table~\ref{tab:visual} and Table~\ref{tab:audio_visual} that scoring approaches based on MS and WATE shows the best results in our experiments. But we would like to note WATE scoring method depends on threshold's selection of recognizability score. Test embeddings with  the recognizability score less than threshold value wasn't used to compute weighted average test embedding for several protocol trial. We chose the threshold value (0.65) using NIST SRE 2021 (development set). JANUS Multimedia Dataset is not suitable to chose the threshold value, because it differs in terms of conditions from NIST SRE 2021 development and evaluation set. Therefore the results in Table~\ref{tab:visual} and Table~\ref{tab:audio_visual} for WATE scoring method are overestimated.

%\begin{table}[]
%\centering
%\caption{Fixed and open video track single systems}
%\label{tab:video}
%\begin{tabular}{|l|c|c|c|}
%\hline
%\rowcolor[HTML]{FFFFC7} 
%\textbf{Fixed track single system} & \textbf{EER} & \textbf{minDCF} & %\textbf{actDCF} \\ \hline
%ResNet101-IR-SE-AN + MS & 0.17  & 0.011 & 0.013 \\ \hline
%ResNet101-IR-SE-AN + WATE & 0.22 & 0.010 & 0.013 \\ \hline 
%ResNet100-PM + MS& 1.32 & 0.013 & 0.018 \\ \hline
% \end{tabular}
%
%\end{table}

\section{Final submitted systems}
\label{sec:submission}

We tried Bosaris Toolkit \cite{BV2013} and  greedy fusion algorithm  \cite{greed_fusion} as a way to combine several single systems into one final submission system. Bosaris toolkit was also used for calibration. All calibration and fusion was trained on the SRE 21 dev sets.
Table \ref{tab:submission} reveals the results of all submitted systems on the dev set.

%Calibration on the pooled dataset not only helps to obtain better results for each task individually but also allows to use one system in both cases.

Greedy fusion is based on iterative process where the most promising system in terms of actDCF is selected to be added on each step until the mean value of actDCF on the dev set stopped improving.

Fused systems produced by Bosaris Toolkit, on the other hand, were difficult to interpret and often included quite similar systems in favor of more diverse ones
Moreover, having such a big list of systems to fuse we see that almost half of them obtain negative weights in the final fusion system.

Given all this, we decided to use following procedure for the final system preparation:
\begin{enumerate}
    \item Calibration model training using \uppercase{Bosaris} Toolkit.
    \item One modality fusion training (audio and video separately) using greedy algorithm.
    \item Fusion of audio and video systems from step 2 for audio-visual track using \uppercase{Bosaris} Toolkit.
    \item Final system score calibration by \uppercase{Bosaris} Toolkit.
\end{enumerate}

\section{Acknowledgments}
\label{sec:foot}
We express our gratitude and deep appreciation to our colleagures from Automatic Speech Recognition Team: Ivan Medennikov, Maxim Korenevsky, Yuri Khokhlov, Mariya Korenevskaya, Tatiana Prisyach and Tatiana Timofeeva for the valuable advises and interesting discussions on the whole duration of the NIST SRE challenge.

%\vfill
%\pagebreak

%\vfill\pagebreak

\bibliographystyle{IEEEbib}

\end{document}